\documentstyle[preprint,aps,eqsecnum]{revtex}
\begin{document}
\tightenlines
\preprint{To Appear in Phys. Rev. D}
\title{
The self-force on a static scalar test-charge outside a 
\\ Schwarzschild black hole}
\author{Alan G. Wiseman}
\address{Enrico Fermi Institute,
University of Chicago\\
5640 Ellis  Avenue Chicago, Illinois 60637-1433\\
{\rm E-mail: agw@gravity.phys.uwm.edu}}
\date{January 11, 2000} 
\maketitle 
\begin{abstract}

The finite part of the self-force on a static scalar test-charge
outside a Schwarzschild black hole is zero.  By direct construction of
Hadamard's elementary solution, we obtain a closed-form expression for
the minimally coupled scalar field produced by a test-charge held
fixed in Schwarzschild spacetime.  Using the closed-form expression,
we compute the necessary external force required to hold the charge
stationary.  Although the energy associated with the scalar field
contributes to the renormalized mass of the particle (and thereby its
weight), {\it we find there is no additional self-force acting on the
charge}.  This result is unlike the analogous electrostatic result,
where, after a similar mass renormalization, there remains a finite
repulsive self-force acting on a static electric test-charge outside a
Schwarzschild black hole.  We confirm our force calculation using
Carter's mass-variation theorem for black holes.  The primary
motivation for this calculation is to develop techniques and formalism
for computing all forces --- dissipative and non-dissipative --- acting
on charges and masses moving in a black-hole spacetime.  In the
Appendix we recap the derivation of the closed-form electrostatic
potential.  We also show how the closed-form expressions for the
fields are related to the infinite series solutions.

\end{abstract}
\bigskip
\pacs{PACS Numbers: 04.30.Nr, 04.70.Bw, 97.60.Lf}

\widetext
\section{\bf INTRODUCTION}
\label{sec: introduction}

In order to gain deeper quantitative understanding of highly
relativistic binary star systems spiraling toward coalescence, a
number of authors have used perturbation theory to study the motion of
test-particles orbiting black holes.  (See {\it e.g.}
\cite{poisson,poissonsasaki,tagoshi,leonardpoisson}.) The basic idea
is to solve the linearized field equations with out-going radiation
boundary conditions on a Schwarzschild or Kerr black-hole background.
The source of the perturbing field is generally chosen to be a test-particle 
(endowed with a small scalar charge, electric charge, or mass
charge) moving on a bound geodesic orbit, {\it e.g.}, a circle.  Once
the perturbing field (scalar, electromagnetic, or gravitational) is
calculated, the energy  radiated per orbit ({\it i.e.,} the
time-averaged energy flux) can be computed by a performing  a surface
integral over a distant sphere surrounding the system.  The rate
energy is carried away by the radiation is then equated to a loss of
orbital energy of the particle; thus one can compute the rate of
orbital decay.\footnote{ Finding a gravitational-wave signal in noisy
detector data requires an accurate prediction of the orbital decay
rate.  See Thorne \cite{kipsnowmass}.} In effect, this energy-balance
argument gives the {\it time-averaged} radiation-reaction force
associated with imposing  out-going radiation boundary conditions
\cite{downgoing}.  This method has been used successfully to compute
the inspiral rate of coalescing binaries to very high relativistic
order, and it has been used to check analogous post-Newtonian
calculation of the inspiral \cite{willwiseman}.  These perturbation
calculations also add insight into effects such as wave ``tails''.
(See \cite{poisson} and \cite{mytailpaper} for post-Newtonian and
perturbation-theory discussions of tails.) However, in spite of the
success of the perturbation approach, the method, as it has been
applied, has a drawback: it has only been used to compute {\it
time-averaged}, {\it dissipative} forces on the particle.

The time-averaging is problematic for two reasons: (1) During the late
stages of inspiral the orbit will be decaying swiftly, and there
will not
be many orbits left before the final splat to average over.  (2) In
the curved spacetime near a black hole, the self-forces experienced by
the particle are not instantaneous forces.  Rather these forces arise
because the fields produced by the particle at one instant travel away
from the particle, encounter the curvature of the spacetime, and then
scatter back and interact with the particle at a later time.  (See
DeWitt and DeWitt  \cite{dewittdewitt} for an illuminating discussion
of this point.) Thus the fields produced when the particle moves
through, say, periastron, and suffers the maximum coordinate
acceleration, will not come back to interact with the particle until
later in the orbit when the coordinate acceleration will be more
gentle.  In other words, the state of motion that produced the fields,
will not be the state of motion affected by the fields.  Although it
is unlikely that taking into account the non-local nature of the
self-forces will qualitatively change any of the results found by the
``averaging'' technique, the issue has never been rigorously addressed
with perturbation theory.

In addition to the dissipative forces, the perturbing fields will
produce other non-dissipative --- conservative  --- self-forces on the
particle, and these forces have not been studied by the conventional
perturbation-theory approach
\cite{poisson,poissonsasaki,tagoshi,leonardpoisson}.  [As DeWitt and
DeWitt \cite{dewittdewitt} have demonstrated, these forces can be
computed using a different perturbative approach.] For example,
suppose we have an electric charge in a circular orbit around a
Schwarzschild black hole.  Even if we neglect the dissipative effects
of the radiation reaction, which cause a secular decay of the orbit,
the charged particle still does not travel on an exact geodesic of the
spacetime.  The particle will feel an additional conservative force ---
proportional to square of the charge --- that pushes it slightly off
the geodesic.  As this force is not radiative in nature, the portion
of the field that produces it is explicitly discarded in the
conventional perturbation approach because it falls off faster than
$O[1/r]$, and therefore any effect it might have on the motion is
missed by a technique that uses a surface integral on a distant sphere
to determine the back-reaction effect.  (2) Another perturbation
calculation by Gal'tsov  \cite{galtsov} has used the
``half-retarded--minus--half-advanced'' Green's function to find the
radiation reaction forces. But these calculations also miss the
conservative ({\it i.e., time symmetric}) parts of the force, because
they simply cancel when the two parts of the Green's function are
subtracted.  In other words, the Gal'tsov result differs from the
DeWitt-DeWitt \cite{dewittdewitt} result in that it misses the
conservative part of
the force.

We can try (but as we will see, fail) to understand the origin of
the conservative forces by considering a static
electric test-charge \cite{testcharge} held fixed
by some non-conducting mechanical struts outside a Schwarzschild
black hole.
The charge  experiences  a self-force (which is clearly not a
radiation-reaction force in any conventional sense) 
proportional to the square of the electric charge.
A physical origin of such a force might be naively described as
follows: the surface of the uncharged black hole will
act as a conductor, and therefore the presence of the external
charge will induce a dipolar charge distribution on the horizon
\cite{conductor}.
The magnitude of the induced dipole moment will depend on the
size of the dipole (in this case, the mass $M$ of the Schwarzschild
black hole) and the magnitude $e$  of the external charge.
This leads one to believe that the charge will experience
an attractive dipolar force scaling as $b_s^{-5}$,
where $b_s$ is the Schwarzschild radial coordinate of the charge.
This force will be in addition to the natural attractive
force (the weight of the particle) which, 
according to Newton, should  scale as $b_s^{-2}$.
Thus our naive suspicion is that, in addition to the weight,
the strut holding the charge fixed will have to counteract 
an attractive self-force of the form
\begin{equation}
F_{\rm (naive)} \sim  { e^2 M \over b_s^5 } \; . \;\;
\label{fnaive}
\end{equation}

Smith and Will (SW) \cite{smithwill} have calculated the
force required to hold an electric charge fixed
outside a Schwarzschild black hole.  They find
the total force exerted by the mechanical strut holding the charge
fixed must be
\begin{eqnarray}
F_{\rm (strut)}={ M \mu_{\rm ren} \over b_s^2} 
\biggl ( 1 - {2 M \over b_s} \biggl )^{-1/2}
- { e^2 M \over b_s^3 } \; . \;\;
\label{electricforce}
\end{eqnarray}
Here $\mu_{\rm ren}$ is the renormalized mass of the test-charge.  We
use units in which $G=c=1$.  As expected, the first term scales as
$O[b_s^{-2}]$, and  shows that the strut must support the weight of
the particle; such a term would be present whether or not the particle
is electrically charged. The second term shows --- contrary to the
``physical intuition'' presented above  --- that the black hole tries
to {\it repel} the charged particle and the repulsion scales as
$O[b_s^{-3}]$.  
\footnote{
Although the second term in Eq.~(\ref{electricforce}) appears only to
be accurate to leading order in $M$, the result is exact for static
charges; the appearance is only a beautiful artifact of Schwarzschild
coordinates.  Switching to harmonic or isotropic coordinates
\cite{coordinates,mtw}  makes the second term appear as an infinite series
in $M$.  See Eq.~(\ref{accelparts}b).
} 
SW show for almost any choice of variables in Eq.~(\ref{electricforce})
the first term dominates, and the second term only lightens the load a
small amount.

The fact that our naive physical intuition led to both the wrong
direction and the wrong scaling of the force suggests such
calculations should be carried out in detail before conclusions are
drawn.  In this paper we carry out the analogous calculation for a
scalar charge in the presence of a Schwarzschild black hole, {\it
i.e.} we compute the force required to hold a {\it scalar} charge
fixed.  The primary result of this paper is to show that, although the
scalar field does contribute to the renormalized mass, there is no
force analogous to the second term in Eq.~(\ref{electricforce}) for a
scalar charged particle.  The primary motivation for this calculation
is much broader: to begin to develop techniques and formalism for
tackling a sequence of problems, namely, calculating all the forces,
linear in the fields (dissipative and non-dissipative), that act on
test-charges and masses moving in the proximity of a black hole.

As the implicit underlying motivation for this work is calculating
the forces on {\it moving} charges, we can ask:
does our no--self-force--on--a--{\it static}--scalar--charge  result
carry over to the case of a moving charge?
Not directly. However,  we can gain some intuition
about extending our static results to dynamic results 
for scalar charges by 
examining the static and dynamic cases for electric charges.
Although the SW force calculation is only valid in the
static limit, DeWitt and DeWitt \cite{dewittdewitt} 
show that when the electric charge is in slow motion
around the black hole, in addition to the conventional radiation reaction force,
there is a conservative repulsive force acting on the charge.
This force is independent of the velocity, and clearly
corresponds to the force found in the static calculation,
{\it i.e.,} the second term in Eq.~(\ref{electricforce}).\footnote{
DeWitt and DeWitt only compute the force to leading 
order in the mass of the hole and 
their result appears to agree 
exactly with the second term in Eq.~(\ref{electricforce}).
However, the agreement is really only at leading order in $M$.
See the previous footnote.} 
Therefore, it seems reasonable to expect a correspondence between
the static case and the slow-motion case for scalar charges.
For scalar charges, this means the absence of a
self-force on a static charge should imply 
there is no conservative (repulsive or attractive) force on 
a slowly orbiting charge.  
This, however, does not rule out the presence of a 
higher-order, conservative
force that scales as, say, the {\it square} of the orbital velocity.
(This issue is under vigorous investigation \cite{tedalan}.)
As a consequence, although both electric and scalar charges
will spiral inward due to radiation reaction forces,
scalar charges will not suffer the same persistent, velocity-independent, 
conservative force trying to hold them off the geodesic.

As a matter of principle, many of the delicate issues of radiation
reaction forces in curved spacetime we will be dealing with in this
paper have already been addressed in quite a general context by
DeWitt and Brehme \cite{dewittbrehme} (using a world-tube
calculation), by Quinn and Wald \cite{quinnwald} (using an axiomatic
approach), and Mino, Sasaki and Tanaka \cite{mino} (using matched
asymptotic expansions).  However, in practice, the results presented
in these papers have never been applied to give even simple results
like Eq.~(\ref{electricforce}), nor did they work out the general
expressions for forces on scalar charges.

We begin in Sec.~\ref{sec: scalarfield} 
by calculating the scalar field produced by
a static scalar charge in Schwarzschild spacetime.
The standard method for solving the field equations
on a black-hole background is to use separation of
variables and decompose the solution into
Fourier, radial and angular modes (spherical harmonics).
(See Appendix~\ref{sec: closedform} for an example.)
The result is an infinite-series solution for the field.
This method works well for computing the far-zone  properties of
the field ({\it e.g.}, the energy flux) where the field is 
weak and can be accurately described by the first few terms in
the series.
However, when computing a force, such as Eq.~(\ref{electricforce})
(or its scalar analogue, as we are doing here),
the behavior of the field on a distant sphere will not suffice;
we need to know the detailed behavior of the field up close to the particle.
For this it will not be sufficient
to describe the field by a few multipoles.
Describing the singular nature of the field near the
charge requires an infinite number of multipoles
\cite{multipoles,jackson};
therefore it is not surprising that SW did not
use the standard multipolar expansion
of the electrostatic field ({\it e.g.}, \cite{cohenwald,hanniruffini})
in their force calculation.
Rather, when embarking on their electrostatic force calculation SW comment:
``... an exact calculation is made possible  by the fortuitous
existence of a previously discovered analytic solution
to the [static] curved-space Maxwell's equations''.
The exact solution to which they refer is an
old result due to
Copson
and modified by Linet \cite{linet} which gives a closed-form
expression for the electrostatic potential of a fixed charge residing in
Schwarzschild spacetime. 
[See Eq.~(\ref{A0sum}).]
The Copson-Linet formula has the advantage that the singular 
nature of the field near the electric charge is manifest,
thus allowing simple calculations in the close proximity of the particle.
We begin by constructing the scalar-field analogue of the 
Copson-Linet result:
a closed-form expression for the field of a fixed 
scalar charge in Schwarzschild spacetime.
Our derivation is similar to Copson's, and is
based on constructing the Hadamard ``elementary solution''
\cite{hadamard}
of the scalar-static field equation.
As in the  Copson-Linet electrostatic solution, 
our solution will clearly show the divergent behavior of the 
field near the particle.
In Appendix~\ref{sec: recapes}, we outline Copson's derivation 
of the closed-form
expression for the electrostatic field with the Hadamard formalism. 
In Appendix~\ref{sec: closedform},
we generate some interesting
summation formulas by equating the closed-form solutions
with the infinite series solutions \cite{cohenwald,hanniruffini}.

As a historical note, we mention that 
Copson \cite{copson} obtained his solution for a static electric charge in 
a Schwarzschild spacetime in 1928. This is approximately 40 years before
the term ``black hole'' was coined \cite{wheeler} 
and researchers began formulating the {\it no-hair} theorem.
However, Copson was probably the first to note one of the important 
features of the {\it no-hair} theorem:
``... the potential of an electron on the boundary
sphere $R=\alpha$ [on the horizon]
is independent of its position on the sphere, a rather curious result.''

Our force calculation in Sec.~\ref{sec: force} closely follows the
SW calculation of the electrostatic force. 
We use the exact expression for the scalar-static field found in
Sec.~\ref{sec: scalarfield} to compute the stress-energy tensor 
and the force density in a local, freely-falling frame near the
fixed scalar charge. 
By integrating the force density 
over a small spherical volume of radius $\bar \epsilon$ centered on the charge
(spherical in the freely-falling frame) we compute
the total force the strut must supply to hold the charge fixed. 
In the limit where the radius of the spherical region of integration shrinks
to zero,
there remains a formally divergent piece of the force scaling 
as $({\rm charge})^2/\bar \epsilon$.
This factor multiplies an acceleration that is
identical to the acceleration in the first term in Eq.~(\ref{electricforce});
therefore the infinite term is simply absorbed into the definition of the mass
of the particle, {\it i.e.,} in Eq.~(\ref{massren}) we define
\begin{equation}
\mu_{\rm ren} \equiv  \mu_{\rm bare} 
+ {1 \over 2} \lim_{\bar \epsilon \rightarrow 0} {q^2 \over \bar \epsilon}  \; ,
\label{renorm}
\end{equation}
where $q$ is the scalar charge of the particle,
and $\mu_{\rm bare}$ is the bare mass of the particle.
In this way, the  scalar charge does contribute to the renormalized mass of the
particle in exactly the same way as the electric charge contributes
to the renormalized mass in Eq.~(\ref{electricforce}).
The form of classical mass renormalization 
depicted in Eq.~(\ref{renorm}) is 
seen in all calculations of this type 
({\it e.g.,}\cite{smithwill,dirac,dewittbrehme}).
The main conclusion of this paper is that,
after the formally infinite piece is absorbed into the renormalized
mass, there remains no scalar counterpart to the
second term in Eq.~(\ref{electricforce}).
Also in Sec.~\ref{sec: force}, we verify this answer
using conservation of energy.

Neither the charge's contribution to the renormalized-mass 
term or to the repulsive term in Eq.~(\ref{electricforce}) 
should be confused with the contribution
to the gravitational force on the particle due the stress energy
of the electric field perturbing the metric of the spacetime.
Such a force would scale as $\mu e^2$.
In this calculation we are explicitly ignoring corrections to the metric.  
See the discussion following Eq.~(\ref{accelparts}e).

In Sec.~\ref{sec: discussion}, we discuss a number of
alternative methods for solving similar problems,
as well as similarities and differences
in the scalar-static and electrostatic results.
We also collect what is known
about the forces on static charges (mass, electric and scalar)
in Schwarzschild spacetime.
This gives a clear indication of what future research is needed.

\section{\bf Solution of the scalar field for a fixed charge in Schwarzschild spacetime}
\label{sec: scalarfield}

In order to find the self-force acting on the scalar charge,
we first solve the massless scalar field equation
\begin{equation}
\Box V 
\equiv (1/ \sqrt{-g}) ( \sqrt{-g} g^{\alpha \beta} V_{,\alpha})_{,\beta}
= 4 \pi \rho
\label{waveeqn}
\;\;
\end{equation}
in Schwarzschild spacetime.\footnote{Since the Ricci scalar 
curvature $R$ vanishes in the 
Schwarzschild spacetime, it would seem that including coupling to
the curvature (e.g. a conformally invariant term ${1\over 6} R V$)
in Eq.~(\ref{waveeqn}) would have no effect on our results. However, if
we include coupling to the curvature, the stress-energy that enters
the force calculation [Eq.~(\ref{sfstressgen})] would have to be modified also.
Therefore, we make no claim that our results hold for non minimally
coupled fields.
}
Here commas denote partial differentiation,
Greek indices run $0$ to $3$, and Latin indices will run $1$ to $3$.
The source $\rho$ for our field will be a point-like \cite{diracdelta}
scalar charge 
which can be described by
\begin{eqnarray}
\rho(t,{\bf x}) &=&  q \int_{-\infty}^{\infty} (1 / \sqrt{-g} ) \;
\delta^4 ( x^{\alpha} - b^{\alpha}(\tau) ) d\tau \nonumber\\
&=& {q \over u^t(t) }  { \delta^3( {\bf x} - {\bf b}(t)) \over \sqrt{-g} } \;,
\label{dynamicsource}
\;\;\;
\end{eqnarray}
where $b^{\alpha}(\tau)$ is the spacetime trajectory of the
charged body, $\tau$ is the proper time measured along the path,
and $u^t =  d b^0 / d \tau =  d t / d \tau $.
[Notice,
$\int \rho d({\rm proper \;\; volume}) = \int \rho u^t \sqrt{-g} d^3 x =q$
in a frame comoving with the charge.]
We will later restrict our attention to a static field and 
a fixed source charge at ${\bf b} = b{\bf \hat z}$,
but, for the present, we will 
leave the time-dependence in the equations.

We use isotropic coordinates \cite{coordinates} to describe the Schwarzschild
geometry.  The line element is
\begin{equation}
ds^2 = - { (2r-M)^2  \over (2r+M)^2 } dt^2 
+ \biggr ( 1 + {M \over 2 r} \biggr )^4  (dx^2 + dy^2 + dz^2 )
\; . \;\;
\label{metric}
\end{equation}
In these coordinates Eq.~(\ref{waveeqn}) can be written
\begin{equation}
C^{ij} \tilde V_{,ij}(\omega,{\bf x}) 
+ C^j \tilde V_{,j}(\omega,{\bf x}) + C \tilde V(\omega,{\bf x}) = 
4 \pi (1 + M/2r)^4 \tilde \rho(\omega,{\bf x}) \; ,
\label{helmholtz}
\;\;\;\;
\end{equation}
where
\begin{mathletters}
\begin{eqnarray}
C^{ij} &=& {\rm diag}(1,1,1) \; , \\
C^j    &=& h(r) { x^j \over r} = {d \over dr} \left[ 
\ln (1-(M/2r)^2) \right ] {x^j \over r} \; , \\
C &=& \omega^2 { (1+M/2r)^6 \over (1- M/2r)^2 } \;,
\end{eqnarray}
\label{cs}
\end{mathletters}
and we have used the Fourier transform
\begin{equation}
V(t,{\bf x}) = \int_{-\infty}^{\infty} \tilde V(\omega,{\bf x})
e^{- i \omega t } d \omega \;
\end{equation}
to eliminate the time variable.
We can also write Eq.~(\ref{helmholtz}) in the more compact form
\begin{eqnarray}
\;\;\;\;\;\;\;\;\;\;\;\;\;\;\;\;\;\;\;\;\;\;\;\;\;
\nabla^2 \tilde V + h(r) \tilde V_{,r} + C(r,\omega) \tilde V
= 4 \pi (1 + M/2r)^4 \tilde \rho(\omega,{\bf x}) \; ,
\;\;\;\;\;\;\;\;\;\;\;\;\;\;\;\;\;\;\;\;\;\;\;\;\;
(\ref{helmholtz}^\prime)
\nonumber
\end{eqnarray}
where $\nabla^2$ is the flat-space Laplacian.
Our primary attention will be focused on the 
static --- zero frequency --- case
($\omega=C=0$), but the formalism we are developing
is valid for the ``Helmholtz''-type equation in
Eqs.~(\ref{helmholtz}) and (\ref{helmholtz}$^\prime$).
Following {\it Hadamard} (p. 92-107),
the elementary solution ({\it i.e.} the Green's function
\cite{laplaceexample}) for Eq.~(\ref{helmholtz})
takes the form
\begin{equation}
U_{elem} = \Gamma^{-1/2}(U_0 + U_1 \Gamma + U_2 \Gamma^2 + \; \dots )  \; ,
\label{hadamardseries}
\;\;\;
\end{equation}
where $\Gamma$ is the square of the geodesic distance (in the sense of the
``metric'' $C^{ij} = \delta^{ij}$)
from the source point ${\bf x^\prime}$ to the field point ${\bf x}$, 
{\it i.e.,}
\begin{equation}
\Gamma = (x-x^\prime)^2 + (y-y^\prime)^2 + (z-z^\prime)^2 \; .
\label{Gamma}
\;\;\;
\end{equation}
Since we are working in three dimensions (an odd number of dimensions),
there is no natural-log term $(\ln \Gamma)$ in the elementary solution.
The $U_n$'s are non-singular functions everywhere outside the
horizon. 
Recall, in isotropic coordinates \cite{coordinates} 
the horizon is at $r=M/2$.
The simple form of $\Gamma$ in Eq.~(\ref{Gamma}) 
is a consequence of the isotropic
coordinates we are using; therefore the technique we are developing
cannot be easily extended
to geometries that do not have spatially isotropic coordinates,
{\it e.g.} Kerr geometry.

The formula for the leading order behavior of the series is given
by {\it Hadamard} (p. 94)
\begin{equation}
U_0({\bf x},{\bf x^\prime})
= {1 \over \sqrt{ \det C^{ij}} } \exp \left \{ - \int_0^\lambda 
\left [ C^{ij} \Gamma_{,ij} +  C^{j} \Gamma_{,j} - 6 \right ] 
{\ d \lambda \over 4 \lambda} \right \} \; ,
\label{U0general}
\;\;\;
\end{equation}
where $\lambda$ is the arc length measured along the geodesic connecting
the source and field points.
Letting $\theta$ denote the angle between the two spatial vectors
${\bf x}$ and ${\bf x^\prime}$ and using Eqs~.(\ref{cs}) and (\ref{Gamma}) 
we have
\begin{mathletters}
\begin{eqnarray}
U_0({\bf x},{\bf x^{\prime}})
&=& \exp \left \{ - \int_0^\lambda \left [ h(r) 
(r-r^\prime \cos \theta) 
\right ] {\ d \lambda \over 2 \lambda} \right \} \; , \\
&=& \exp  \left \{ - \int_{r^\prime}^r  
h(r^{\prime \prime}) { dr^{\prime\prime} \over 2 } \right \} \;,\\
&=& \sqrt{ {1-(M/2r^\prime)^2 \over 1-(M/2r)^2} }  \; , 
\label{U0}
\;\;\;
\end{eqnarray}
\end{mathletters}
where $r=|{\bf x}|$, $r^\prime =|{\bf x^\prime }|$,
and $r^{\prime \prime}$ is the dummy integration variable over $r$.
We have also used the geometric relationship 
$(r-r^\prime \cos \theta) d\lambda = \lambda dr$.

Because we are looking for an axially symmetric solution,
we may assume
the $U_n$'s are functions of only the radial variables
[{\it e.g.,} notice $U_0 = U_0(r,r^\prime)$].
We now substitute 
Eq.~(\ref{hadamardseries}) into Eq.~(\ref{helmholtz}),
use relations such as $\delta^{ij} \Gamma_{,i}  \Gamma_{,j} = 4 \Gamma$,
and collect powers of $\Gamma$.  The result is
\begin{eqnarray}
&& -(2U_{0,r} + h U_0) \left ( { r^2-{r^\prime}^2 \over 2 r } \right )
\Gamma^{-3/2} 
+ \sum_{n=0}  \biggl [(2n+1)(2U_{n+1,r} + h U_{n+1} )  
\biggl ({ r^2-{r^\prime}^2 \over 2 r } \biggr )  \nonumber \\
&&
+ 2 (2n^2+n-1) U_{n+1} 
+{1 \over r} (r U_n)_{, r r} + h U_{n,r} 
+ { 2n -1 \over 2 r} ( 2U_{n,r} + h U_n) + C(r,\omega) U_n  \biggr ]
\Gamma^{n-1/2} = 0 . \nonumber \\
\label{recur1}
\end{eqnarray}
Since $U_{elem}$  is a solution to the homogeneous equation  
everywhere (except at the source point) the 
coefficients of each power of  $\Gamma$  must vanish independently. 
The first term gives an equation for $U_0$
\begin{equation}
2U_{0,r} + h U_0 = 0 \; .
\label{hdef}
\;\;\;
\end{equation}
Notice that our Eq.~(\ref{U0general}) has already given us
a particular solution (Eq.~(\ref{U0})) to this equation.
Setting the coefficient of $\Gamma^{n-1/2}$ to zero and using
an integrating factor, we obtain a recursion relation
for the $U_n$'s
\begin{eqnarray}
{U_{n+1}(r,r^\prime) \over U_0} 
= 
{-1 \over (r^2 - {r^\prime}^2)^{n+1} } \int_{r^\prime}^r &&
{ ({r^{\prime \prime}}^2 - {r^\prime}^2 )^n \over 2n+1 }
\biggl [ { ( r^{\prime \prime} U_n)_{, r^{\prime \prime}r^{\prime \prime}} 
 \over U_0 } 
- { 2 r^{\prime \prime} U_{0,r^{\prime \prime} }
 U_{n, r^{\prime \prime}r^{\prime \prime}}  \over U_0^2 }
\nonumber \\
&&
+ (2n-1) (U_n/U_0)_{, r^{\prime \prime}} 
+ r^{\prime \prime}C(r^{\prime \prime},\omega) 
(U_n/U_0) \biggr ] dr^{\prime \prime}
\; .
\label{intrecurs}
\end{eqnarray}
In the integrand the $U_n$'s are functions of the dummy
integration variable $r^{\prime \prime}$ and the 
source point $r^{\prime}$, {\it i.e.} 
$U_n=U_n(r^{\prime \prime},r^\prime)$.
This recursion relation allows us to construct
a series solution for the Green's function for the
Schwarzschild-Helmholtz equation Eq.~(\ref{helmholtz}) 
or (\ref{helmholtz}$^\prime$).\footnote{
We can gain confidence in the recursion relation
Eq.~(\ref{intrecurs}) by applying it to
the true Helmholtz
equation in flat spacetime (Eq.~(\ref{helmholtz}) with $M=0$, but
$C\equiv \omega^2\neq 0$).  In this case, Eq.~(\ref{U0general})
gives $U_0=1$, and the recursion relation Eq.~(\ref{intrecurs})
gives $U_1 = -\omega^2/2$, $U_2 = \omega^4/24$, $\dots$,
$U_n = (-1)^n \omega^{2n}/(2n!)$.
Summing the series gives  the  Green's function for the
Helmholtz equation
\begin{eqnarray}
U_{elem} = { \cos(\omega |{\bf x} - {\bf x^{\prime}} | ) \over
| {\bf x} - {\bf x^{\prime}}| }  \; . \nonumber
\end{eqnarray}
Now compute the inverse Fourier transform and we
have 
\begin{eqnarray}
G(x,x^\prime) = {1\over 8\pi } \biggl \{
{\delta[t^\prime-(t-|{\bf x}-{\bf x^\prime}|)]\over |{\bf x}-{\bf x^\prime}|}
+
{\delta[t^\prime-(t+|{\bf x}-{\bf x^\prime}|)]\over |{\bf x}-{\bf x^\prime}|}
\biggr \}  \; . \nonumber
\end{eqnarray}
This is the half-advanced {\it plus} half-retarded  Green's function.
See \cite{jackson}, Eq.~(6.61).
}

We now explicitly assume that the charge and the field are static,
{\it i.e.} we assume that $\omega=C(r,\omega)=0$ in
Eq.~(\ref{intrecurs}).
Beginning with $U_0$ from Eq.~(\ref{U0}) we can construct
\begin{mathletters}
\begin{eqnarray}
{U_1 \over U_0 } &=& 
- {1 \over 2} { (2M)^2 \over (4{r^\prime}^2 - M^2) (4r^2 - M^2) }   
\equiv - {1 \over 2} \gamma  \; , \\
{U_2 \over U_0 } &=& 
\;\;\; 
{3 \over 8} { (2M)^4 \over (4{r^\prime}^2 - M^2)^2 (4r^2 - M^2)^2 }  
=  {3 \over 8} \gamma \; , \\
{U_3 \over U_0 } &=& 
-{5 \over 16} { (2M)^6 \over (4{r^\prime}^2 - M^2)^3 (4r^2 - M^2)^3 }  
=  - {5 \over 16} \gamma \; .
\end{eqnarray}
\label{calus}
\end{mathletters}
Substituting these into Eq.~(\ref{hadamardseries}),
the pattern is immediately clear and the summation is elementary
\begin{mathletters}
\begin{eqnarray}
U_{elem}({\bf x}, {\bf x^\prime}) &=& {U_0 \over \sqrt{\Gamma} } 
\biggl ( 1 - {1 \over 2} \gamma \Gamma 
           + {3 \over 8} (\gamma \Gamma )^2
           - {5 \over 16} (\gamma \Gamma )^3  + \dots \biggr )  \\
         &=& {U_0 \over \sqrt{\Gamma} } 
             {1 \over \sqrt{ 1 + \Gamma \gamma } } \; 
              \label{uelemsum} \;\;\;
              \\
&=& {1 \over \sqrt{\Gamma} }
{ r (4 {r^\prime}^2 - M^2) \over r^\prime
\sqrt{ (4 {r^\prime}^2 - M^2) (4 r^2 - M^2) + 4 M^2 \Gamma } } \; .
\label{uelem}
\;\;
\end{eqnarray}
\end{mathletters}
Equation~(\ref{uelem}) is a closed-form expression for the
Green's function for Eq.~(\ref{helmholtz}) (with C=0).
In the limit $M=0$ it reduces to  $|{\bf x} - {\bf x^{\prime}}|^{-1}$,
the Green's function for Poisson's equation in flat space.
We use Eq.~(\ref{uelem}) in precisely the same manner we
would use the Green's function for Poisson's equation:
we integrate it against the source to obtain a particular solution
to the inhomogeneous equation.
In our case the source is the static scalar test-charge
held fixed at the spatial position ${\bf b}$, and we have
\begin{mathletters}
\begin{eqnarray}
V_{part}({\bf x},{\bf b}) 
&=& - \int 
\biggl [ {q \over u^t(b) \sqrt{-g(b)} } \biggl ( 1 + {M \over 2 b } \biggr )^4
\delta^3 ({\bf x^\prime}-{\bf b}) \biggr ] U_{elem}({\bf x},{\bf x^\prime})
 d^3 {\bf x^\prime} \\
&=& - q  { 2b-M \over 2b+M }  { 1 \over \sqrt{\Gamma ({\bf x},{\bf b})} } 
{ 4 r b \over \sqrt{ (4 b^2 - M^2) (4 r^2 - M^2) + 4 M^2 
\Gamma ({\bf x},{\bf b}) } } \\
&=& - q  \sqrt{ {b_h-M \over b_h+M } } { 1 \over 
\sqrt{ r_h^2 - 2 r_h b_h \cos \theta + b_h^2 - M^2 \sin^2 \theta } } \; .
\end{eqnarray}
\label{partic}
\end{mathletters}
In the last two steps we have explicitly included the factor 
$1/u^t(b) = \sqrt{-g_{00}(b)} = (2b - M) /(2b + M ) =
\sqrt{ (b_h-M )/( b_h+M ) }$; 
in the last step we have converted to harmonic coordinates 
\cite{coordinates}. 
The leading minus sign in Eq.~(\ref{partic}) is a consequence
of the source term in Eq.~(\ref{waveeqn}); our source is
`$+$'$4\pi \rho$ and not the familiar `$-$'$4\pi \rho$
of electrostatics.
(Like scalar charges attract.)
As we are now dealing strictly with a
static solution, we have also dropped the {\it twiddle} denoting the
Fourier transform in Eq.~(\ref{helmholtz}).

Equation~(\ref{partic}) is a {\it particular} solution to the scalar-static
field equation, but is it the desired solution to the
equation?  
In other words, does it satisfy all the boundary conditions?
First, the field and its derivatives are well behaved
outside --- and on --- the horizon;
thus our solution has no unphysical regions of infinite energy
(save, of course, at the location of the charge).
For $r \gg b > M/2 $ we see from Eq.~(\ref{partic}b)
\begin{equation}
V_{part}(r \rightarrow \infty ) 
\approx  - {q \over r}  \biggl ( { 2b-M \over 2b+M } \biggr ).
\label{asympbehavior}
\;\;
\end{equation}
As the charge is lowered toward the horizon, the factor
$(2b-M)/(2b+M )$ extinguishes
the field measured by a distant observer.
Notice if the charge is lowered to the horizon ($b = M/2$) the
field completely disappears.
(See, {\it e.g.}, \cite{teitelboim} for discussion.)
Thus, it is the extinction factor --- which had its origin
in  the factor of $1/u^t(t)$ in Eq.~(\ref{dynamicsource})
and has survived throughout the calculation ---
that enforces {\it no scalar hair on the black hole}.\footnote{
The factor $1/u^t$ is present even in a special relativity.
Notice it is just the inverse of the Lorentz factor
$\gamma = dt/d\tau = 1/\sqrt{1-v^2}$.
As a mnemonic, the strength of the ``charge'' depends on the spin of the field:
for masses (spin 2) we have $m = \gamma^1 m_{\rm rest}$,
for electric charges  (spin 1) we have $e = \gamma^0 e_{\rm rest}$,
and clearly, by induction, for scalar charges
(spin 0) we have $q = \gamma^{-1} q_{\rm rest}$.
}

Equation~(\ref{asympbehavior}) is the appropriate asymptotic 
form of the scalar field; therefore 
the particular solution Eq.~(\ref{partic})
is the desired solution which satisfies all the 
boundary conditions \cite{boundryconditions,teukolsky}
\begin{equation}
V({\bf x},{\bf b}) = V_{part}({\bf x},{\bf b}) \; .
\end{equation}

Before embarking on the force calculation, we note
a remarkable feature of the closed-form solution 
we have found.  Despite the fact that every term in the
Hadamard series is divergent at the horizon [Eq.~(\ref{calus})],
the closed-form expression for the elementary solution
is well behaved on the horizon.
The easiest way to see this is with the harmonic-coordinate expression
Eq.~(\ref{partic}c) evaluated on the horizon ($r_h=M$)
\begin{equation}
V_{Horizon} = - {q \over b_h - M \cos \theta } \sqrt{ {b_h-M \over b_h+M} } \;.
\end{equation}
It is also interesting to note that the horizon is not 
a surface of constant ``potential''.  
In Sec.~\ref{sec: introduction}
our (flawed) intuition about the self-force on a static electric charge
was predicated on the horizon acting as conducting surface;
therefore
we should not be surprised that the force on the static scalar 
charge will be different.

\section{\bf Self-force on a static scalar charge}
\label{sec: force}
\subsection{local method}
\label{subsec: local}

We begin our calculation of the self-force with a picturesque
description ({\it Gedankenexperiment}) of how
such a measurement  could be made.
We imagine a test-charge with bare mass $\mu_{bare}$ and 
scalar charge $q$ held fixed by a non-conducting system of
mechanical struts outside the horizon of a Schwarzschild black hole.
A non-conducting experimenter
at the apex of a vertical, ballistic trajectory
momentarily comes to rest with respect to the fixed charge.
At this moment, she reaches out and measures the force required to hold the
charge fixed, {\it i.e.} she measures the force needed to just lift the
charge off the strut.
The spacetime event $\cal{B}$ where/when the force is measured will be
taken as the origin of the free-falling
observer's coordinates $x^{\bar \alpha}$.
(The over-bar denotes coordinates in the local, freely-falling frame
of the observer \cite{coordinates}.)
Clearly, by symmetry, we can choose our coordinate system such that
the particle is located on the $z$-axis, and the freely-falling coordinate
system is aligned so that the $\bar z$-axis coincides with the $z$-axis.
Thus the only component of
the force for the freely-falling observer to measure will
be $F^{\bar z}$. 
Although this is an elaborate scheme to define the force measurement,
working in the freely-falling frame where the
charge is momentarily at rest is the surest way to 
establish unambiguously
how the scalar field of the particle contributes
to the renormalized mass.

As no scalar charges have
ever been observed, nor scalar fields measured,
our assumption that the scalar field does not interact
with the experimental apparatus (the struts) and the experimenter
seems to be quite plausible.

In order to compute the force in the 
free-falling frame of the observer, we need to be able
to convert quantities from the isotropic coordinates of
Eq.~(\ref{partic}) to the coordinates of the
observer.
The defining feature of the free-falling frame is that spacetime 
is locally flat: {\it i.e.},
\begin{mathletters}
\begin{eqnarray}
&& (g_{\bar \alpha \bar \beta})_{\cal{B}} = {\rm diag} (-1,1,1,1) \equiv 
\eta_{\bar \alpha \bar \beta} \\
&& (g_{\bar \alpha \bar \beta, \bar \gamma})_{\cal{B}} =0  \\
&& g_{\bar \alpha \bar \beta}(x^{\bar \gamma}) = \eta_{\bar \alpha \bar \beta}
+ O[(x^{\bar \gamma})^2] \; .
\end{eqnarray}
\label{freefallmetric}
\end{mathletters}
In particular, the Christoffel symbols,
$\Gamma^{\bar \alpha}_{\bar \beta \bar \gamma}
= (1/2)g^{\bar \alpha \bar \sigma}[g_{\bar \sigma \bar \gamma,\bar \beta}
+ g_{\bar \sigma \bar \beta,\bar \gamma } - g_{\bar \beta  \bar \gamma ,\bar \sigma }]
\sim O[\bar x^{\bar \sigma}]$ near the event $\cal{B}$ .
Fortunately, SW
have done the dirty work in finding the transformations
from isotropic coordinates (un-barred) to the free-falling coordinates 
(barred). They are given by
\begin{mathletters}
\begin{eqnarray}
\bar t &=& {1-M/2b \over 1+M/2b} t + {M \over b^2 }
{ 1 \over (1+ m/2b)^2 } t (z-b) + O[(x^\mu -b^\mu)^3] \\
x^{\bar j} &=& (1+M/2b)^2 (x^j-b^j) + {M \over 2b^2 } {1-M/2b \over (1+M/2b)^5}
\delta^{\bar j \bar 3} t^2 \nonumber \\
&& \;  - {M \over 2b^2 }(1+ m/2b)
[2(x^j-b^j)(z-b) - \delta^{\bar j \bar 3}|{\bf x} - {\bf b}|^2 \, ]
+ O[(x^\mu -b^\mu)^3] \; .
\end{eqnarray}
These can be used to find
\begin{eqnarray}
{1 \over \sqrt{ \Gamma({\bf x},{\bf b})} } =
{1 \over |{\bf x} - {\bf b}| } = 
{1 \over |{\bf \bar x} - {\bf \bar X}(\bar t)|} 
(1+M/2b)^2 \biggl [ 1 - {M \over 2 b^2 }
{1 \over (1+M/2b)^3} \bar z + O[(x^{\bar \alpha})^2] \biggr ] \; ,
\end{eqnarray}
where ${\bf \bar X}(\bar t)$ is the position of the charge as viewed
in the freely-falling frame
\begin{equation}
X^{\bar j} = 
{ 1 \over 2} a_g \delta^{\bar 3 \bar j} \bar t^2 +O[\bar t^3] \; ,
\label{traj}
\;\;
\end{equation}
and $a_g$ is the acceleration of the fixed charge
as measured in the freely falling frame 
at the moment the experimenter comes to rest
at the apex of her geodesic trajectory
\begin{eqnarray}
a_g && \equiv {M \over b^2 } {1 \over (1+M/2b)^3 (1-M/2b) } \; 
\label{ag}
\\
&& ={ M \over b_s^2} 
\biggl ( 1 - {2 M \over b_s} \biggl )^{-1/2} \;\; .
\end{eqnarray}
\label{transformation}
\end{mathletters}
In the second line we have converted to Schwarzschild coordinates.
Notice this is the same as the acceleration appearing
in Eq.~(\ref{electricforce}).
We can use Eq.~(\ref{transformation}) to evaluate the scalar field
and its derivatives in the free-falling frame
\begin{mathletters}
\begin{eqnarray}
&&V(\bar t,{\bf \bar x}) = 
- { q \over |{\bf \bar x} - {\bf \bar X}(\bar t)| }
\biggl [ 1 - {1\over 2} a_g \bar z + O[(x^{\bar \alpha})^2] \biggr ]  \; ,\\
&&V(\bar t=0,{\bf \bar x}) = 
- q \biggl [ {1\over \bar r } - {a_g \over 2 }n^{\bar 3}  
+ O[\bar r \times {\rm even \;number\;of\;} (n^{\bar l}){\rm \,'s} ]  \biggr ]
 \; , \\
&&V_{,\bar t} (\bar t=0,{\bf \bar x}) = - q \, 
O[\bar r^0 \times {\rm odd\;number\;of\;}(n^{\bar l}){\rm \,'s}]  \; , \\
&&V_{,\bar k} (\bar t=0,{\bf \bar x}) 
= - q \biggl [ - {n^{\bar k} \over {\bar r}^2 }  
+ {a_g \over 2 \bar r } (n^{\bar k} n^{\bar 3} - \delta^{\bar 3 \bar k} )
+ n^{\bar k}  O[\bar r^0 \times 
{\rm even\;number\;of\;} (n^{\bar l}){\rm \,'s} ] \biggr ] \; , \\
&&V_{,\bar t \bar t}(\bar t=0,{\bf \bar x}) =
 - q \biggl [  {a_g \over {\bar r}^2 }  
- {a_g^2 \over 2} {n^{\bar 3} \over \bar r} 
+ O[\bar r^{-1} \times {\rm even\;number\;of\;} (n^{\bar l}){\rm \,'s}
] \biggr ]  \; , \\
&&V_{,\bar k \bar t}(\bar t=0,{\bf \bar x}) =
- { q \over \bar r } \biggl [ 
n^{\bar k} O[\bar r^0 \times {\rm odd\;number\;of\;}
(n^{\bar l}) {\rm \,'s}] \biggr ] \; ,
\end{eqnarray}
\label{fieldt0}
\end{mathletters}
where $\bar r = |{\bf \bar x}|$ and $n^{\bar k} = x^{\bar k}/\bar r$.
The spatial indices can be freely raised and lowered with $\delta^{ij}$.

Notice the extinction factor present in Eq.~(\ref{partic}) 
does not extinguish the charge in the freely-falling 
frame Eq.~(\ref{fieldt0}a), that is,
in the freely-falling frame the dominant behavior of the
field is simply $({\rm charge}/{\rm distance})$, independent of how
deep the charge is in the Schwarzschild potential.

We will compute the force required to hold the charge fixed
by integrating the force density \cite{smithwill,weinberg}
\begin{equation}
f^{\bar z} = T^{\bar z \bar \beta } _{\;\;\;\;\; ; \bar \beta}
\label{forcedensity}
\;\;
\end{equation}
over the physical extent of the charged body 
at the instant of time ($\bar t = 0$) when the measurement is made.
More precisely, since we are describing the particle as 
a Dirac $\delta$-function, we will integrate over an infinitesimal sphere of
radius $\bar \epsilon$ centered on the particle and take the
limit as $\bar \epsilon \rightarrow 0$.
The stress-energy tensor $T^{\bar z \bar \beta }$ will
have contributions from the bare mass of the particle and
the scalar field; thus we have
\begin{mathletters}
\begin{eqnarray}
F_{\rm (strut)}^{\bar z} &=& \lim_{\bar \epsilon \rightarrow 0}
\int_{ \bar r \leq \bar \epsilon} 
[ f^{\bar z} ]_{\bar t = 0} \; d^3 \bar x 
= \lim_{\bar \epsilon \rightarrow 0}\int_{ \bar r \leq \bar \epsilon}
\biggl [ T^{\bar z \bar \beta } _{\;\;\;\;\; ; \bar \beta}  \biggr ]_{\bar t = 0}
\; d^3 \bar x  \\
&=& \lim_{\bar \epsilon \rightarrow 0}\int_{ \bar r \leq \bar \epsilon}
\biggl [ T^{\bar z \bar \beta } _{{\rm (bare)} \; ; \bar \beta}
\biggr ]_{\bar t = 0}
\; d^3 \bar x
\; + \;
\lim_{\bar \epsilon \rightarrow 0}\int_{ \bar r \leq \bar \epsilon}
\biggl [ T^{\bar z \bar \beta } _{{\rm (SF)} \; ; \bar \beta} 
\biggr ]_{\bar t = 0} \; d^3 \bar x \;. 
\label{forceintegral}
\;\;
\end{eqnarray}
\end{mathletters}
The first term in the Eq.~(\ref{forceintegral}) shows that the
strut must support the bare weight of the particle.  This
term is present whether or not the particle is charged.
Using the stress-energy tensor for the bare mass of a point particle
located at ${\bf b} = b{\bf \hat z}$
\begin{equation}
T_{\rm (bare)}^{\alpha \beta} = \mu_{\rm bare} 
{ \dot b^{\alpha}(b) \dot b^{\beta}(b) \over \sqrt{-g(b)} u^t(b) } 
\delta^3({\bf x} - {\bf b})  \; , \;\;
\label{stressbare}
\end{equation}
SW found the necessary force the strut must supply to
support the bare weight of the particle to be
\begin{equation}
F_{\rm (bare)}^{\bar z} = 
 \lim_{\bar \epsilon \rightarrow 0}\int_{ \bar r \leq \bar \epsilon}
\biggl [ T^{\bar z \bar \beta } _{{\rm (bare)} \; ; \bar \beta}
\biggr ]_{\bar t = 0}
\; d^3 \bar x
= {M \mu_{\rm bare} \over b^2 } {1 \over (1+M/2b)^3 } {1 \over 1-M/2b }
= \mu_{\rm bare} a_g \; .
\label{fbare}
\;\;
\end{equation}
The second term in Eq.~(\ref{forceintegral}) involves the
stress-energy tensor of the scalar field.
In evaluating this integral we make use  of Eq.~(\ref{freefallmetric})
and note that the connection coefficients in this frame
are $O[x^{\bar \alpha}]$; thus we can write
\begin{eqnarray}
F_{\rm (SF)}^{\bar z} && =
\lim_{\bar \epsilon \rightarrow 0}
\int_{ \bar r \leq \bar \epsilon}
\biggl [ T^{\bar z \bar \beta }_{{\rm (SF)} \; ; \bar \beta}
\biggr ]_{\bar t = 0} \; d^3 \bar x \nonumber \\
&& =
\lim_{\bar \epsilon \rightarrow 0}  \biggl \{
\int_{ \bar r \leq \bar \epsilon} 
\biggl [ T^{\bar z \bar k}_{{\rm (SF)} \; , \bar k}
+T^{\bar z \bar t} _{{\rm (SF)} \; , \bar t}
+O[x^{\bar \alpha} T^{\bar \beta \bar \gamma}]
\biggr ]_{\bar t = 0} \; d^3 \bar x
\biggr \} \; .
\label{sfforce}
\end{eqnarray}
We will denote the three contributions to 
Eq.~(\ref{sfforce}) as $F_{\rm (SF1)}^{\bar z}$, $F_{\rm (SF2)}^{\bar z}$
and $F_{\rm (SF3)}^{\bar z}$ respectively.
We also make use of Eq.~(\ref{freefallmetric})
in writing the stress-energy tensor for the scalar field
\begin{mathletters}
\begin{eqnarray}
T^{\bar \beta \bar \gamma} 
&\equiv& 
{1\over 4 \pi} \biggl [ g^{\bar \alpha \bar \sigma} g^{\bar \beta \bar \tau}
V_{,\bar \sigma} V_{,\bar \tau} 
- {1\over 2}  g^{\bar \alpha \bar \beta} g^{\bar \sigma \bar \tau}
V_{,\bar \sigma} V_{,\bar \tau}
\biggr ] 
\label{sfstressgen}
\;\;
\\
\label{sfstress}
&=& {1\over 4 \pi} \biggl [ \eta^{\bar \alpha \bar \sigma} 
\eta^{\bar \beta \bar \tau}
V_{,\bar \sigma} V_{,\bar \tau} 
- {1\over 2}  \eta^{\bar \alpha \bar \beta} \eta^{\bar \sigma \bar \tau}
V_{,\bar \sigma} V_{,\bar \tau} 
+ O[(x^{\bar \alpha})^2 V_{,\bar \sigma} V_{,\bar \tau} ]
\biggr ] \; .
\;\;
\end{eqnarray}
\end{mathletters}
We now substitute Eq.~(\ref{sfstress}) into Eq.~(\ref{sfforce})
and treat the terms in reverse order.
The third term in Eq.~(\ref{sfforce}) gives a contribution of the form
\begin{eqnarray}
F_{\rm (SF3)}^{\bar z} &&= 
{1\over 4 \pi} \lim_{\bar \epsilon \rightarrow 0} 
\int_{ \bar r \leq \bar \epsilon}
\biggl [
O[x^{\bar \alpha} T^{\bar z \bar t}_{\rm (SF)} ] 
\biggr ]_{\bar t = 0} d^3 \bar x \nonumber \\
&& =  {1\over 4 \pi} \lim_{\bar \epsilon \rightarrow 0}
\int_{ \bar r \leq \bar \epsilon}
\biggl [
O[x^{\bar \alpha} V_{,\bar \beta} V_{,\bar \gamma} ]
+ O[(x^{\bar \alpha})^3 V_{,\bar \beta} V_{,\bar \gamma}]
\biggr ]_{\bar t = 0} d^3 \bar x  \; .
\end{eqnarray}
Using Eq.~(\ref{fieldt0}), we see the most singular
terms come from $(V_{,\bar k} V_{,\bar l})$; therefore
we have
\begin{eqnarray}
F_{\rm (SF3)}^{\bar z} \sim \lim_{\bar \epsilon \rightarrow 0}
\int_{ \bar r \leq \bar \epsilon}
\biggl [ && {1\over \bar r } \,\times \,
[{\rm term\;with\;odd\;number\;of\;} (n^{\bar l}){\rm 's}] 
\nonumber \\
&& +\, {\bar r }^0 \, \times \,
[{\rm term\;with\;even\;number\;of\;} (n^{\bar l}){\rm 's} ] \biggr ]
d \bar r d \bar \Omega \; .
\end{eqnarray}
The first term contains an odd number of unit vectors,
and therefore will vanish when we integrate 
over the solid angle.  
The second term vanishes as ${\bar \epsilon \rightarrow 0}$,
and similarly for higher powers of $\bar r$.
These two tricks are used repeatedly in evaluating the remaining
integrals in Eq.~(\ref{sfforce}).
Thus we have 
\begin{equation}
F_{\rm (SF3)} = 0 \; .
\label{fsf3} \;\;
\end{equation}

We now evaluate the second term in Eq.~(\ref{sfforce}) using
Eq.~(\ref{fieldt0}) and ruthlessly discarding terms that do not 
survive the limit
or the angular integration:
\begin{mathletters}
\begin{eqnarray}
F_{\rm (SF2)}^{\bar z} &&= - \lim_{\bar \epsilon \rightarrow 0}
{1\over 4 \pi} \int_{ \bar r \leq \bar \epsilon}
\left [ V_{,\bar z} V_{,\bar z \bar t} \right ]
{\bar r}^2 d \bar r d \bar \Omega \\
&& =
{1 \over 4 \pi} q^2 a_g
\lim_{\bar \epsilon \rightarrow 0} \int_0^{\bar \epsilon} {1 \over \bar r^2}
d\bar r
\oint n^{\bar z}  n^{\bar z} d \bar \Omega \\
&& =
{1 \over 3} q^2 a_g \lim_{\bar \epsilon \rightarrow 0} 
\int_0^{\bar \epsilon}   {1 \over \bar r^2} d\bar r \; . \;\;
\label{fsf2}
\end{eqnarray}
\end{mathletters}

Using the divergence theorem and  Eq.~(\ref{sfstress}),
the first term in Eq.~(\ref{sfforce}) can be written
\begin{eqnarray}
F_{\rm (SF1)}^{\bar z} &&=  {1\over 4 \pi}\lim_{\bar \epsilon \rightarrow 0}
\int_{ \bar r \leq \bar \epsilon} 
\biggl [  T^{\bar z \bar k}_{{\rm (SF)} \, , \bar k}
\biggr ]_{\bar t = 0} d^3 \bar x
= \lim_{\bar \epsilon \rightarrow 0}
\oint_{ \bar r = \bar \epsilon} 
\biggl [  T^{\bar z \bar k}_{\rm (SF)} \biggr ]_{\bar t = 0}
n^{\bar k} {\bar r}^2 d \bar \Omega  \nonumber \\
&& =  {1\over 4 \pi} \lim_{\bar \epsilon \rightarrow 0}
\oint_{ \bar r = \bar \epsilon}
\biggl [ V_{,\bar z} V_{,\bar k}
- {1\over2} \delta^{\bar z \bar k} 
( - V_{,\bar t} V_{,\bar t} + \delta^{\bar l \bar m} V_{,\bar l} V_{,\bar m} )
+ O[(x^{\bar \alpha})^2 V_{,\bar \beta} V_{,\bar \gamma} ]
\biggr ]_{\bar t = 0} {\bar r}^2 n^{\bar k} d \bar \Omega \; . \;\;
\label{sf1}
\end{eqnarray}
All the terms except the first vanish either by
$\oint \dots d\bar \Omega =0$, 
or $\lim_{\bar \epsilon \rightarrow 0} \dots =0$.
Using Eq.~(\ref{fieldt0}) and  integrating the first term in
Eq.~(\ref{sf1}) over the solid angle,  we are left with
\begin{equation}
F_{\rm (SF1)}^{\bar z} =
{1 \over 3 } a_g \lim_{\bar \epsilon \rightarrow 0}
{q^2 \over \bar \epsilon }
\; . \;\;
\label{fsf1}
\end{equation}

Combining the results in Eqs.~(\ref{forceintegral}), (\ref{fbare}),
(\ref{fsf3}), (\ref{fsf2}) and (\ref{fsf1}) we 
have
\begin{mathletters}
\begin{eqnarray}
F_{\rm (strut)}^{\bar z} &&=
\biggl [ \mu_{\rm bare} + 
{q^ 2 \over 3} \lim_{\bar \epsilon \rightarrow 0} 
\left ( {1 \over \bar \epsilon } +  
\int_0^{\bar \epsilon}   {1 \over \bar r^2} d\bar r \right ) \biggr ] a_g \\
&& =
\biggl [ \mu_{\rm bare} + {q^2 \over 3} \int_0^\infty    {1 \over \bar r^2} 
d \bar r
\biggr ] a_g \\
&& = \mu_{\rm ren } a_g \; ,
\label{finalforce}
\;\;
\end{eqnarray}
\end{mathletters}
where we
have defined the leading factor in the bracket
as the ``renormalized'' mass
$\mu_{ren}$.
The renormalized mass has the same functional form as the renormalized
mass for the electric charge in the SW calculation.
Converting to Schwarzschild coordinates \cite{coordinates}, we get
\begin{eqnarray}
F_{\rm (strut)}={ M \mu_{\rm ren} \over b_s^2} 
\biggl ( 1 - {2 M \over b_s} \biggl )^{-1/2} 
\; + \left \{ {\rm \; nothing \; depending \;on\;} q \; \right \}
\; .
\label{forceSchw}
\;\;
\end{eqnarray}
{\it  There is 
no self-force} of the form seen in the second term of Eq.~(\ref{electricforce}).

\subsection{global method}

We verify our no-self-force result
by means of a global, energy-conservation calculation.
Suppose, instead of measuring the force on the charge
while the charge is in place (as we did in the last subsection),
the free-falling observer lowers the charge
a small amount $\delta \bar b$.
The work done on the experimenter will be
\begin{equation}
\delta \bar W = - F^{\bar z} \delta \bar b = - F^{\bar z} (1+M/2b)^2 \delta b 
\; ,
\end{equation}
where we have used Eq.~(\ref{metric}) to convert the 
free-falling displacement $\delta \bar b$ to an isotropic
coordinate displacement $\delta b$.
The experimenter then converts this energy into
a photon and fires the photon to asymptotic infinity.
The energy received at infinity will be red-shifted
\begin{equation}
\delta E_{\rm received} = \sqrt{ -g_{00}(b) } \; \delta \bar W \; .
\end{equation}
By conservation of energy this change in the system will 
be manifested by a change in asymptotic mass ${\cal M}$ of the system
\begin{eqnarray}
\delta {\cal M} &&=  - \delta  E_{\rm received} 
= [1-(M/2b)^2]  F^{\bar z} \delta b \; .
\end{eqnarray}
Thus we have 
\begin{equation}
 F^{\bar z} = { 1 \over 1 - (M/2b)^2 } { \delta {\cal M} \over \delta b } \; .
\label{forcemass}
\;\;
\end{equation}

We now use the {\it total mass variation law}  of Carter \cite{carter},
which shows how the asymptotic mass will differ between two
situations where the gravitational and matter status of the spacetime
is slightly altered. 
In our case we compute the difference in asymptotic mass
before and after we make the small displacement
of the charge.
The relationship is given by
\begin{equation}
\delta {\cal M}  - { \kappa \over 8 \pi } \delta {\cal A} 
= { 1 \over 8 \pi } \delta \int G^0_{\;\;\; 0} \sqrt{ - g } d^3 x
+ { 1 \over 16 \pi } \int G^{\mu \nu} h_{\mu \nu} \sqrt{ - g } d^3 x
\; .
\label{carter}
\;\;
\end{equation}
Here $\kappa$ is the surface gravity of the black hole, ${\cal A}$ is the area
of black hole, $G^{\mu \nu}$ is the Einstein tensor, and
$h_{\mu \nu}$ is the difference in the metric between  the
two configurations.
In Eq.~(\ref{carter}) we have neglected terms involving the spin of
the black hole.
Throughout this paper we have assumed that
the metric is unperturbed by the 
presence of the charge; therefore the last term in Eq.~(\ref{carter})
vanishes.
For a Schwarzschild black hole the {\it area} term  in Eq.~(\ref{carter})
is just  the change in the mass of the black hole.
During our slow displacement we will assume that no matter or
radiation goes down the hole, and therefore this term will vanish.
(We  revisit this point at the end of the section.)
Using Einstein's equation to write $G^0_{\;\; 0} = 8\pi \; T^0_{\;\; 0}$
all that remains of Eq.~(\ref{carter}) is
\begin{eqnarray}
\delta {\cal M}  &&= \delta \int T^0_{\;\;\; 0} \;\; \sqrt{ - g }  \; \; d^3 x 
\nonumber \\
&&= \delta \int T^0_{{\rm (bare)}  0} \; \; \sqrt{ - g } \;\; d^3 x 
 + \delta \int T^0_{{\rm (SF)}   0} \; \; \sqrt{ - g } \;\; d^3 x  \; .
\label{deltam}
\;\;
\end{eqnarray}
The first integral can be computed by Eq.~(\ref{stressbare}).
SW give the  result
\begin{equation}
{\cal E}_{\rm bare} \equiv \int T^0_{{\rm (bare)}  0} 
\;\;\sqrt{ - g } \;\; d^3 x
= \mu_{\rm bare} \biggl ( {1-M/2b \over  1+M/2b } \biggr ) \; .
\label{ebare}  
\;\; 
\end{equation}
Noting that the metric is diagonal, the scalar field is static, and
employing the definition of the stress tensor for a scalar field
Eq.~(\ref{sfstressgen}), we can write the second integral
in Eq.~(\ref{deltam}) as
\begin{eqnarray}
{\cal E}_{\rm SF} && \equiv \int T^0_{{\rm (SF)}  0} \;\; \sqrt{-g} \;\; d^3 x
\; = \; {1\over 8 \pi} 
\int \biggl [ g^{jk} V_{,j} V_{,k} \biggr ] \sqrt{-g} \;\; d^3 x
\nonumber \\
&& = 
{1\over 8 \pi}  \int \biggl [ 
( g^{jk} V_{,j} V \sqrt{-g} )_{,k}
- V ( g^{jk} V_{,j} \sqrt{-g} )_{,k}
\biggr ] \;\; d^3 x \; .
\label{esf1}
\;\;
\end{eqnarray}
The first term in Eq.~(\ref{esf1}) can be converted to two surface integrals:
one over the horizon, the other over a sphere at $r\rightarrow \infty$.
The field and its derivatives are well behaved on the horizon, 
but $g^{jk} \sqrt{-g}$ 
vanishes there; therefore the integral on the horizon vanishes.
The other surface integral vanishes in the limit $r\rightarrow \infty$.
Using the original field equation Eq.~(\ref{waveeqn}), we have 
\begin{equation}
( \sqrt{-g} g^{jk} V_{,j})_{,k} =  4 \pi \sqrt{-g}\rho \; .
\end{equation}
Thus we can write
\begin{equation}
{\cal E}_{\rm SF} = - {1 \over 2} \int \rho V \sqrt{-g} d^3 x
=  -  {q \over 2 u^t(b)} \int V \delta^3({\bf x} - {\bf b})
d^3 x \; .
\label{esf} \;\;
\end{equation}
Unfortunately, the scalar field is divergent at the source point
{\bf b}; therefore we must renormalize. We do so by modeling our source as a
charged, spherical shell of radius $\epsilon$, {\it i.e.}
\begin{equation}
{ \delta^3({\bf x} - {\bf b})  \over \sqrt{-g} }
\; \rightarrow \;
\lim_{\epsilon \rightarrow 0}
 { \delta^1( |{\bf x} - {\bf b}| - \epsilon ) \over
  4\pi \epsilon^2 } \; .
\end{equation}
Performing the integration and using the metric to  convert the radius
of the ball to free-falling coordinates 
$\bar \epsilon  = (1+M/2b)^2 \epsilon$, we have
\begin{equation}
{\cal E}_{\rm SF} = 
 {1 \over 2} \biggl ( { 1-M/2b \over 1+M/2b } \biggr ) 
\lim_{\bar \epsilon \rightarrow 0} {q^2\over \bar \epsilon}
\; . \;\;
\label{esffinal}
\end{equation}
As expected the functional form is the same here as in
Eq.~(\ref{ebare}). 

Combining Eq.~(\ref{ebare}) and  Eq.~(\ref{esffinal})
defining the renormalized mass
\begin{equation}
\bar \mu_{\rm ren} = \mu_{\rm bare} +
{1\over2} \lim_{\bar \epsilon \rightarrow 0} {q^2\over \bar \epsilon} \; ,
\label{massren}
\end{equation}
and using Eq.~(\ref{forcemass}) and Eq.~(\ref{ag}), we have
\begin{equation}
F^{\bar z} = \bar \mu_{\rm ren} a_g \; .
\end{equation}
This agrees exactly with our previous calculation
Eq.~(\ref{finalforce}): no finite part of the self-force.

We close this section with a pedagogical
comment on the global energy conservation
method for computing the force.
We note that the calculation was predicated on the assumption
that the {\it area} term in Eq.~(\ref{carter}) vanished,
{\it i.e.},
\begin{equation}
{\kappa \over 8 \pi } \delta {\cal A} = 
\delta ({\rm mass \;\; of \;\; the \;\; hole }) = 0 \; .
\end{equation}
In terms of modern black hole theory, this is a
valid assumption: No particles were dropped into the hole.
The charge was displaced slowly,
so no transverse fields ({\it i.e.} radiation) heated the horizon.
Therefore, the area remains unchanged.
However,
in Sec.~\ref{subsec: local} we 
gave a primitive derivation of the force which 
did not appear to explicitly rely on any sophisticated properties of 
black holes.
(Implicitly, we did assume that the mass of the hole remained constant
when the observer wiggled the  charge to measure the force.)
An interesting interpretation of the
two force calculations is to accept the primitive
derivation in Sec.~\ref{subsec: local} as the correct force.
Then, when we evaluate the right hand side of 
Eq.~(\ref{deltam}) and show that it is the same 
as our local  (primitive) force calculation,
we have {\it verified} that the area of
the black hole did not change when we lowered the charge.
This tells us, in spite of the fact that energy of the scalar-static
field (or the electric field for an electric charge) 
extends clear down to the
horizon, that none of the energy near the horizon is 
pushed across the horizon when we lower the charge.

\section{\bf Conclusions  and Discussion}
\label{sec: discussion}

We have developed a formalism for constructing the 
Hadamard elementary solution for the Schwarzschild-Helmholtz
equation (\ref{helmholtz}). 
This formalism was chosen because it 
expresses the singular nature of field
in the near proximity of the charge.
In the case of a static charge, we are able to 
find a closed-form expression for the field.
We have used this expression to show (after mass renormalization)
there is no self-force
on a static scalar charge outside a Schwarzschild black hole

Although we have patterned our discussion after Copson \cite{copson}
and SW \cite{smithwill},
there are alternative methods for computing the scalar field
and the self-force.
For example, in the static limit, Linet \cite{linetgreensfunction}
has used {\it generalized axially symmetric potential theory} (GASP)
\cite{weinstein} to derive the Green's function for the scalar
field.
However, this type of construction has not been
extended to the Helmholtz-type equation depicted in Eq.~(\ref{helmholtz}).
Lohiya \cite{lohiya} has demonstrated a concise method for
determining the force on a static electric charge.
Lohiya's method also uses the Copson-Linet closed-form expression 
for the electrostatic potential;
however it is unclear how to extend Lohiya's method to moving particles.
The formalism we have developed can be extended to moving
charges.  The recursion relation, Eq.~(\ref{intrecurs}), can
be integrated with $\omega \ne 0$.
Although it may be hard to find a simple summation
of the results as in Eq.~(\ref{uelemsum}),
it is possible to obtain the Green's function
to the first several orders in $\Gamma$. 

Now that we have computed the (absence of) forces acting on a static scalar charge,
let us assemble what is known about all the forces
on a static charge outside a Schwarzschild black hole.
In order to express this, let us slightly change the thought experiment. 
We will give the test-charge a mass $\mu$, an electric
charge $e$, and a scalar charge $q$.
We will support the charge on a strut as before,
but, instead of measuring the force supplied by the strut at some moment
$\bar t = 0$, we kick the strut out from under the charge and find
the instantaneous acceleration of the falling charge in harmonic coordinates. 
After the particle begins to move there will also be
a radiation-reaction force, so we must make the measurement at the moment
we remove the strut.
The metric can be used to convert quantities from free-falling
(proper) coordinates to harmonic coordinates \cite{coordinates}.
The result is
\begin{equation}
\biggl [ \mu {d^2 r_h \over d t_h^2 } \biggr ]_{\bar t=0}
 = \biggl ( {r_h-m \over r_h+m} \biggr )^{3/2}
F^{\bar z}_{\rm (strut)}
\; .
\end{equation}
Using this to convert Eqs.~(\ref{electricforce}) and (\ref{forceSchw}),
and expanding in the post-Newtonian quantity $M/r_h$, we have
\begin{mathletters}
\begin{eqnarray}
\biggl [ \mu {d^2 r_h \over d t_h^2 }   \biggr ]_{\bar t = 0}
= {M\over r_h^2}
\biggl \{ && \;\;\; \mu \biggl [  -1 + 4 \biggl ( {M\over r_h} \biggr ) 
- 9 \biggl ( {M\over r_h} \biggr )^2
+ 16 \biggl ( {M\over r_h} \biggr )^3
\; + \;\; {\rm known \;\; terms }   \;\; \biggr ]   \\
&& + {e^2 \over M} \biggl [   \biggr ({M\over r_h}  \biggl )
- 6 \biggl ( {M\over r_h} \biggr)^2  + {39\over 2}  
\biggl ( {M\over r_h} \biggr)^3 
\; + \;\;   {\rm known \;\; terms } \;\;  \biggr ]   \\
&& + {q^2 \over M} \biggl [  
{\rm \; \; all \;\; terms \;\; are \;\; known \;\; to \;\; be \;\;  zero \;\; }
\biggr ]   \\
&& + {\mu^2 \over M} \biggl [  2  \biggr ( {M\over r_h} \biggl )
- {87 \over 4} \biggl ( {M\over r_h} \biggr)^2 \; +\;\;   
{\rm unknown \;\; terms } \;\; 
\biggr ]  \\
&& + \; O[e^2 \mu] \; +\;  O[ q^2 \mu ] \; +\;  O[\mu^3 ]
\biggr \} \; .
\end{eqnarray}
\label{accelparts}
\end{mathletters}
Here $r_h$ denotes the radial position of the particle in 
harmonic coordinates.
In line (\ref{accelparts}a) we have recovered the velocity 
independent terms of the geodesic equation of motion expressed in 
harmonic coordinates. Since this is just a Taylor expansion of the
first term in Eq.~(\ref{electricforce}), we know
these terms to all orders in  $M/r_h$.
Thankfully, the first three terms are in agreement
with the second post-Newtonian equations of motions.
(See {\it e.g.} \cite{damour300,lincolnwill}.)
Line (\ref{accelparts}b) is just the
expansion of the second term in  Eq.~(\ref{electricforce}),
and thus we know these terms to all orders in $M/r_h$.
Line (\ref{accelparts}c) is the scalar-charge part, which we
have shown to vanish for all orders in $M/r_h$.
For moving charges, there will very likely be non-zero terms.
In line (\ref{accelparts}d), only the first two
terms are known from second post-Newtonian calculations.
(See {\it e.g.} \cite{damour300,lincolnwill}.)
The question remains, can the unknown
terms in  line (\ref{accelparts}d) be obtained by methods
similar to those used to find the electric and scalar forces,
that is, by looking at the field (metric) perturbations produced by the
mass of the test-particle?
Obviously there are a number of conceptual issues to tackle
in answering this question.
For example, when the metric itself is the perturbed field,
can we define a freely falling observer in the 
same way as we did in the scalar-charge force calculation?
When solving for the metric perturbation, how do the stresses
in the strut affect the solution?
This is currently under vigorous investigation \cite{wiseman}.
Line (\ref{accelparts}e) represents higher order effects,
such as additional forces on the particle due to the 
change in the metric produced by the electric field of the particle.
Such terms will also arise from a gauge change, say, 
$r\rightarrow r+O[\mu]$.
These terms are clearly second order in the perturbations.

\acknowledgments

It is a pleasure to thank Patrick Brady, Ted Quinn and Bob Wald
for many useful conversations.
This work was supported 
in part by NASA Grant NAGW-4268 and NSF Grant AST-9417371
while the author was at Caltech, and by NSF Grant PHY95-14726
while the author was at the University of Chicago.



\appendix
\section{\bf Recap of electrostatic charge in Schwarzschild spacetime }
\label{sec: recapes}

In this section we summarize the results of Copson \cite{copson}
and 
obtain the closed-form solution for the electrostatic potential
using our Hadamard construction.
Using isotropic coordinates and assuming the field is strictly static,
Maxwell's equations for the electrostatic potential can be
written
\begin{equation}
C^{ij} A_{0,ij}({\bf x}) 
+ C^j A_{0,j}({\bf x}) = 
-16 \pi e { b^2(2a-M) \over (2b+M)^3 } \delta^3({\bf x} - {\bf b})  \; ,
\label{emstat}
\;\;\;
\end{equation}
where 
\begin{equation}
C^j    = h(r) { x^j \over r} = {d \over dr} \biggl \{ 
\ln \biggr [ { (1+M/2r)^3 \over  1-M/2r } \biggr ] \biggr \} {x^j \over r} \; .
\end{equation}
Equation~(\ref{emstat}) is in the same form as Eq.~(\ref{helmholtz});
therefore we proceed using Eq.~(\ref{U0general}) and we get
\begin{equation}
U^{\rm (el)}_{0} = {r \over r^{\prime} }
\biggl ( {2r^{\prime} + M \over 2r+M } \biggr )^{3/2}
\biggl ( {2r - M \over 2r^{\prime} - M } \biggr )^{1/2} \; .
\end{equation}
The superscript $({\rm el})$ denotes that these are parts of the 
electrostatic solution.
Using the recursion relation Eq.~(\ref{intrecurs}),
we have
\begin{mathletters}
\begin{eqnarray}
U^{\rm (el)}_1(r,r^{\prime}) &&= \;\;\; {3 \over 2} U^{\rm (el)}_0 \gamma 
\; ,\\
U^{\rm (el)}_2(r,r^{\prime}) &&= - {5 \over 8} U^{\rm (el)}_0 \gamma^2   
\; ,\\
U^{\rm (el)}_3(r,r^{\prime}) &&= \;\;\; {7 \over 16} U^{\rm (el)}_0 \gamma^3
\; , \\
U^{\rm (el)}_4(r,r^{\prime}) &&= - {45 \over 128} U^{\rm (el)}_0 \gamma^4 
\; ,
\end{eqnarray}
\label{elus}
\end{mathletters}
where $\gamma$ is the same as in Eq.~(\ref{calus}a). Once
again the summation is elementary:
\begin{equation}
U^{\rm (el)}_{\rm elem}(r,r^{\prime}) = 
{ U^{\rm (el)}_{0} \over \sqrt{ \Gamma} } 
{ (1 + 2 \gamma \Gamma) \over \sqrt{ 1 + \Gamma \gamma } } \; . \;\;
\label{elelem}
\end{equation}
As in the scalar case, we integrate the elementary 
solution against the source and obtain a particular solution
to electrostatic field Eq.~(\ref{emstat})
\begin{equation}
A^{\rm part}_{0}({\bf x},{\bf b})
=   { e \over (b_h+M) (r_h+M) } 
{   b_h r_h  -M^2 \cos \theta   \over 
\sqrt{ r_h^2 - 2 r_h b_h \cos \theta + b_h^2 - M^2 \sin^2 \theta } } \; ,
\label{A0part}
\;\;
\end{equation}
where we have switched to harmonic coordinates \cite{coordinates}.
This is Copson's \cite{copson} 1928 solution ``for the potential
of an electron in the Schwarzschild field''.
As in the scalar case we must ask: does the particular
solution satisfy the boundary conditions? It is well behaved at
the horizon, so there is no problem there. 
However,
as $r_h\rightarrow\infty$  the potential does not give the
correct value
\begin{equation}
A^{\rm part}_{0}(r_h\rightarrow\infty)
\sim { e \over r_h } { b_h \over b_h+M } \neq  { e \over r_h } \; .
\label{asymp}
\end{equation}
The fact that the field does not behave as $e/r_h$ for 
large $r_h$ suggests
by Gauss's law that we have found a solution with some additional charge
lying around. 
However, our solution satisfies the homogeneous (source-free)
equation 
everywhere outside the horizon except at the source point where there
is a charge $e$.
We must conclude that we found a solution with some charge on the
horizon.
In order to fix the boundary condition, we need to 
add a monopolar solution of the homogeneous equation,
that is, we need to add an image charge.
It is easy to see what is needed,
\begin{equation}
A^{\rm homog}_{0}= { e M \over (r_h+M)(b_h+M) } \; ,
\label{A0homog}
\;\;
\end{equation}
and check that this satisfies the homogeneous equation 
outside the horizon.
Linet \cite{linet} noticed the discrepancy in Eq.~(\ref{asymp})
and added this piece to Copson's result.
Combining the two pieces gives the final result: the electrostatic
potential for a fixed charge outside a Schwarzschild black hole
\begin{equation}
A_0({\bf x},{\bf b})
=   { e \over (b_h+M) (r_h+M) }  \left \{
{   b_h r_h  -M^2 \cos \theta   \over 
\sqrt{ r_h^2 - 2 r_h b_h \cos \theta + b_h^2 - M^2 \sin^2 \theta } } 
 + M \right \} \; .
\label{A0}
\;\;
\end{equation}

The potential $A^{\rm part}_{0}$ in Eq.~(\ref{A0part}) is the particular
solution constructed directly from the Hadamard elementary 
solution.
If the  Hadamard potential $A^{\rm part}_{0}$ is taken to be the actual 
potential and the force calculation is carried out
({\i.e.,} repeat the SW calculation using the method similar to
Sec.~\ref{subsec: local}),
the resulting force is zero.
This means that although our construction of the Hadamard solution
was strictly a local calculation, when we summed the series
we found a solution with just enough charge on the
horizon to cancel the repulsive force in Eq.~(\ref{electricforce}).
This also means that the repulsive force that SW found 
for the electric charge is due solely to the part of the potential
$A^{\rm homog}_{0}$ which is tacked on to satisfy the boundary conditions.
In other words, the second term in Eq.~(\ref{electricforce})
is simply the force produced by the image charge on the horizon,
and the force can be computed from Eq.~(\ref{A0homog}) directly
\begin{eqnarray}
F_{\rm self} && = e \biggl [ 
{d \over d r_h } A^{\rm homog}_{0} (r_h, b_h) \biggr ]_{r_h = b_h}  
\nonumber \\
&& = - { e^2 M \over (b_h + M)^3 } = -  { e^2 M \over b_s^3 } \; . \;\;
\label{fimage}
\end{eqnarray}
This gives a physical interpretation of the repulsive force;
the charge outside the black hole is repelled by an image charge
inside the horizon.

\section{\bf Comparing closed-form solutions with series solutions }
\label{sec: closedform}

Equating our closed-form solutions for the scalar-static and electrostatic
fields with the conventional infinite series solutions,
we can obtain some interesting summation formulas
that do not appear in the
standard references \cite{stegun,whittakerwatson,gradshteyn}.

For a fixed point source Eq.~(\ref{waveeqn}) is easily solved by
separation of variables. The angular dependence is expressed
by Legendre polynomials, and the resulting radial equation
is also Legendre's equation. 
Equating the series solution
to our close-form solution Eq.~(\ref{partic}c) we have
\begin{mathletters}
\begin{eqnarray}
V({\bf x_h},&&{\bf b_h})
= -  q\sqrt{ \frac{b_h-M}{b_h+M} }
\frac{1}{ \sqrt{ r_h^2 - 2 r_h b_h \cos \theta + b_h^2 - M^2 \sin^2 \theta }}
\\
&& = - {q \over M} \sqrt{ \frac{b_h-M}{b_h+M} } \sum_{l=0}^{\infty}
( 2 l + 1 ) P_l(\cos \theta )
\left \{
\begin{array}{ll}
   P_l(r_h/M) \, Q_l(b_h/M) \;\; , &  \rm{if} \;\; r_h < b_h  \\
   P_l(b_h/M) \, Q_l(r_h/M) \;\; , &  \rm{if} \;\; r_h > b_h
\end{array}
\right \} .
\end{eqnarray}
\label{modesum}
\end{mathletters}
Here the $P_l$ and $Q_l$ are the Legendre functions.
This summation is a special case of 
Equation 28 in MacRobert \cite{macrobert}.

If the field point is located on the horizon ($M=r_h < b$) 
or on the axis ($\theta=0$) we can verify this formula
using the standard summation formula \cite{stegun,whittakerwatson,gradshteyn}
\begin{equation}
\sum_{n=0}^{\infty} (2n+1)\, Q_n(x) \, P_n(y) = \frac{1}{x-y}
\;\;\;\;\;\;\;\; |x|>1 \; \rm{and} \; |x|>|y| \;.
\label{legendresum}
\end{equation}
Applying this summation  formula to Eq.~(\ref{modesum}),
we get
\begin{equation}
V(\rm{horizon})=-\frac{q}{b_h - M \cos \theta } \sqrt{\frac{b_h-M}{b_h+M} }
\;.
\end{equation}
Clearly the horizon is not a surface of constant ``potential''.
This is in contrast with the electrostatic case where
the horizon is a surface of constant potential.
[See Eq.~(\ref{A0}).]
On the axis of symmetry
({\it i.e.,} $\theta = 0$, so $P_l(cos \theta ) = 1$ for
all values of $l$) we can also use Eq.~(\ref{legendresum}) to sum
Eq.~(\ref{modesum})
\begin{equation}
V(\rm{axis})=-\frac{q}{|r_h -b_h|} \sqrt{\frac{b_h-M}{b_h+M} } \; .
\end{equation}

In the electrostatic case, the series solution can 
be similarly obtained by separation of variables. 
The radial functions are derivatives of Legendre functions.
Equating the analytic expression with the series solution
gives an interesting summation formula
\begin{mathletters}
\begin{eqnarray}
&&A_0({\bf x},{\bf b})
={ e \over (b_h+M) (r_h+M) }
\left \{ { b_h r_h  -M^2 \cos \theta   \over 
\sqrt{ r_h^2 - 2 r_h b_h \cos \theta + b_h^2 - M^2 \sin^2 \theta } } 
 + M \right \} \\
&&= - {e \over M^3}  \sum_{l=0}^{\infty}
{ 2 l + 1 \over  l (l+1) }  P_l(\cos \theta )
\left \{
\begin{array}{ll}
(r_h-M)(b_h-M) P_l^{\prime} (r_h/M) \, Q_l^{\prime} (b_h/M) 
\;\; , &  \rm{if} \;\; r_h < b_h  \\
(r_h-M)(b_h-M) P_l^{\prime} (b_h/M) \, Q_l^{\prime} (r_h/M) 
\;\; , &  \rm{if} \;\; r_h > b_h
\end{array}
\right \} \;.
\nonumber \\
\end{eqnarray}
\label{A0sum}
\end{mathletters}
Here it is understood when $l=0$ we make the replacement
\begin{equation}
{ (r_h-M) \over l}  P_l^{\prime} (r_h/M)  \rightarrow M
\;\;\;\;\; ({\rm when} \;\;\; l=0  ) \; .
\end{equation}
Notice that the horizon ($r_h=M$) is a surface of constant potential.
The closed-form expression [modulo the homogeneous piece in Eq.~(\ref{A0homog})]
was computed in 1928 by Copson \cite{copson},
who expanded the result in terms of
radial functions and discovered the summation formula.
The series result was rederived by Cohen and Wald \cite{cohenwald},
and Hanni and Ruffini \cite{hanniruffini}.
The asymptotic form of the solution can be seen immediately 
from the closed-form result, 
or from the series by 
noting that $Q_0(x\rightarrow\infty) \approx 1/x$. 
We see that
\begin{equation}
A_0(r\rightarrow\infty) \approx {e \over r} \; ,
\end{equation} 
which is the correct behavior.
Equation~(\ref{modesum}) can also be 
obtained from Eq.~(\ref{A0sum}) by differentiating
and using Legendre's equation.


\begin{references}
\bibitem{downgoing}
At high relativistic order it is also necessary to satisfy
a strictly ``in-going'' boundary condition at the horizon.
See \cite{poissonsasaki}.
\bibitem{conductor}
It is clear from Eq.~(\ref{A0sum}) that the horizon $r_h=M$ is
a surface of constant ``potential''; therefore, in this sense (only),
the horizon behaves like a conductor.  Computing the lines of
force gives a different picture.
\bibitem{coordinates}
We use three different types of coordinates to
describe the Schwarzschild geometry: 
isotropic coordinate, Schwarzschild coordinates
and harmonic coordinates. 
The radial coordinates will be
denoted $r$, $r_s$ and  $r_h$ respectively. They are
related to each other by
$r_s = r ( 1 + M/(2r) )^2  = r_h + M$, 
$r = (1/2)\{ r_s - M [r_s(r_s-2M)]^{1/2} \} = (1/2)[r_h + (r_h^2-M^2)^{1/2} ]$,
$r_h = r_s - M = r[ 1+ (M/2r)^2 ]$.
The horizon of the black hole is located at
$r = M/2$, $r_s = 2M$ and  $r_h = M$.
The coordinates with an over-bar refer to coordinates in the
freely-falling frame.  The spacetime event where the
force measurement is made is $(\bar t,\bar x,\bar y,\bar z) =(0,0,0,0)$.
We adopt units in which $G=c=1$.
\bibitem{multipoles}
The simplest example of this is
the field of an isolated electric charge $e$ in flat spacetime where the
field is  V=$e/d$. 
(Here $d$ is the distance from the charge). 
However, if the origin of the coordinates is located somewhere
other than the charge we can also expand
$e/d = 4 \pi e \sum_{(l,m)}^{\infty} (2l+1) (r_<^l/r_>^{(l+1)}) 
Y_{lm}^*(\theta^\prime,\phi^\prime) Y_{lm}(\theta,\phi)$.
(See \cite{jackson} page 102.)
Clearly, a simple expression like $e/d$ is more convenient
than the infinite sum
when we need to the know the behavior near the singularity.
\bibitem{diracdelta}
Our convention for the Dirac-$\delta$ function is
$\delta^4 (x-y)= \delta(x^0-y^0) \delta(x^1-y^1)
\delta(x^2-y^2) \delta(x^3-y^3)$
and $\int \delta^4 (x-y) dx^0 dx^1 dx^2 dx^3 = 1$.
However, we pedantically write the scalar quantities
${ \delta^4 (x-y) \over \sqrt{-g} }$ and  $\sqrt{-g} d^4x$.
\bibitem{boundryconditions}
In general, if a particular solution does not satisfy the
boundary conditions, you patch it up by adding a solution
to the homogeneous equation to fix the boundary values.
However, in the present case it
is a good thing that the particular solution we found satisfies
the boundary conditions, 
because, in Schwarzschild spacetime,
there are no static solutions the  homogeneous scalar field equation
\cite{teukolsky}.
\end{references}
\end{document}